\documentclass{vgtc}                          

\usepackage{mathptmx}
\usepackage{graphicx}
\usepackage{times}
\usepackage{amsmath}
\usepackage{amsfonts}
\usepackage{url}
\usepackage{tikz}
\def\checkmark{\tikz\fill[scale=0.4](0,.35) -- (.25,0) -- (1,.7) -- (.25,.15) -- cycle;} 
\usepackage{multirow}

\onlineid{0}

\vgtccategory{Research}

\vgtcinsertpkg

\title{C-D ratio in multi-display environments}

\author{Travis Gesslein\thanks{e-mail: travis.gesslein@gmail.com}\\ %
        \scriptsize Mixed Reality Labor, Hochschule Coburg %
\and Jens Grubert \thanks{e-mail: jens.grubert@gmail.com}\\ %
     \scriptsize Mixed Reality Labor, Hochschule Coburg }%

\abstract{Research in user interaction with mixed reality environments using multiple displays has become increasingly relevant with the prevalence of mobile devices in everyday life and increased commoditization of large display area technologies using projectors or large displays. Previous work often combines touch-based input with other approaches, such as gesture-based input, to expand the possible interaction space or deal with limitations of other two-dimensional input methods. In contrast to previous methods, we examine the possibilities when the control-display (C-D) ratio is significantly smaller than one and small input movements result in large output movements. To this end one specific multi-display configuration is implemented in the form of a spatial-augmented reality sandbox environment, and used to explore various interaction techniques based on a variety of mobile device touch-based input and optical marker tracking-based finger input. A small pilot study determines the most promising input candidate, which is compared to traditional touch-input based techniques in a user study that tests it for practical relevance. Results and conclusions of the study are presented. } 

\CCScatlist{ 
  \CCScat{H.5.2}{HCI}%
{Multimedia Information Systems}{Input Methods};
  \CCScat{I.3.7}{Computer Graphics}{Three-Dimensional Graphics and Realism}{Virtual reality, augmented reality, mixed reality}
}

\begin{document}


\firstsection{Introduction}

\maketitle


When using any arbitrary input device, some process has to map the performed input onto some output (such as position). This mapping process is not limited to virtual or augmented reality environments, but can be achieved in the real world as well. The relation of how some input is mapped onto output is the C-D ratio:

\begin{equation}
CD_{ratio} = \frac{\Delta x}{\Delta X} \label{eqcdratio}
\end{equation}

It maps a movement distance \(\Delta x\) in physical space to a mapped movement distance \(\Delta X\) in the output space. C-D ratios of one or close to one are commonly used in application such as smartphones where a touch on the display surface is usually mapped directly to, for example, a virtual button press in the same position on the display.
In contrast, applications where small inputs are mapped onto large outputs become relevant when the space being interacted with is larger than the input device. C-D ratios close to one can still be used in these cases, but it is desirable to investigate C-D ratios that directly map some input on the small input device onto a target in the larger display space in one step.
One application field where such problems appear are multi-display environments, where multiple displays can be combined to overcome limitations of single-display technologies. The experimentation setup described here in particular uses an spatial augmented reality environment based on the RoomAlive Toolkit \cite{jones2014roomalive} which combines a projector that projects an image onto some arbitrarily shaped surface with a commodity smartphone that is able to show part of the same projected environment on its screen in higher resolution, while also offering the ability to interact with the space directly using touch controls. In the following we show separate but related simple mathematical models for mapping positional input onto output positions. Several novel input methods that map small input spaces onto large output spaces based on the described mathematical frameworks are presented and implemented. Developed input methods are evaluated and put in contrast with each other in a user study, which offers insight into practical relevance of small C-D ratio mappings. 

\section{Related Work}

\cite{Blanch:2004:SPI:985692.985758} suggests that user performance is dependent on movement in the physical control space (also called motor space), as opposed to output movement that results from the C-D ratio mapping on the target display. This presents an opportunity for improvement for problems such as the occlusion problem of touch interfaces presented in \cite{ryall2006experiences} by decoupling output movement from input movement. \cite{igarashi2000speed} investigates zooming as a potential C-D ratio modifier by introducing a method to automatically scale a document that is read by the user based on the current scrolling speed through the document. The technique is similar to Z-Scaling, presented here later.  One popular mode of 3D input are gesture-based methods. Previous work such as \cite{parker2006integrating} \cite{subramanian2006multi} \cite{hilliges2009interactions} \cite{izadi2008going} \cite{izadi2008going} describe techniques where input of traditional input devices is augmented with hand-gestures that are recognized using the support of added hardware and algorithms. \cite{marquardt2011continuous} introduces the idea of a continuous interaction space, and publications in the following years use this concept, such as\cite{Chen:2014:AIT:2642918.2647392} and \cite{Song:2012:HBM:2207676.2208585}, which seek to unify gesture and touch-based input, or \cite{Han:2015:TCT:2702613.2732849} which expands the touch input area using air above and around a smartphone. Instead of gesture based input, the work in this thesis makes use of positional finger information above a smartphone surface to implement movement akin to mouse movement. 
Much research into 3D target acquisition exists in the fields of virtual and augmented reality. \cite{argelaguet2013survey} surveys more than 30 of these VR and AR object selection techniques and presents some conclusions as to the main concerns and limitations of their design.
In \cite{fitts1954information}, Paul Fitts presents a predictive mathematical model for human movement in target selection, known as Fitts' law, and \cite{Mackenzie:1992:FLP:143848} introduces the now commonly used reformulation of the index of difficulty appearing in Fitts' Law in terms of the Shannon-Hartley theorem \cite{hartley1928transmission} 

\begin{equation}
ID = log_2(\frac{D}{W} + 1) \label{eqfittslaw}
\end{equation}

in order to deal with limitations of the original formulation, where D and W are distance to the target and W its width (or radius), respectively. This thesis does not examine the correctness of results in tests of the presented input methods with respect to Fitts' law, but the definition above is still used as a guideline for the relationship between the size and distance of a target that is shown to the user.
\cite{jones2014roomalive} introduces the RoomAlive toolkit, a software library that enables spatial augmented reality (SAR) applications in projector-camera systems by providing the necessary algorithms to correctly render virtual environments onto arbitrarily shaped surfaces. The toolkit is used in work presented here in order to experiment with and implement various input methods in an SAR environment. \cite{grubert2017headphones} establishes a mobile multi-display environment based on commodity smartphones by tracking the user's head and face and using it as a reference, thus showing mobile possibilities of multi-display environments.
The work presented in this thesis uses optical marker-based tracking in order to track various objects. Previous work such as \cite{macritchie2015integrating} suggest gluing a marker onto a subject's finger nails, joints and other hand positions in order to reconstruct a model of the user's hand skeleton via inverse kinematics. \cite{dorfmuller2001finger} suggests attaching markers to commodity gloves that are then worn by users for hand and finger tracking. However, it is unclear whether gloves are possibly a confounding factor in user performance. \cite{sun2007empirical}, \cite{jokinen2016interactive} and \cite{hwangbo2013study} suggest that gloves have no effect on performance in the particular input methods tested in the respective papers, however, research in other scientific fields such as \cite{tiefenthaler2006touch}, \cite{dianat2012methodology} and \cite{bishu1993investigation} suggest influence of gloves on various hand-related task. To remove these uncertain variables, the work presented here aims at a glove-less tracking solution. 

\section{Concept}
\begin{figure} \label{fig:twocases}
    \centering
    \includegraphics[width=3.0in]{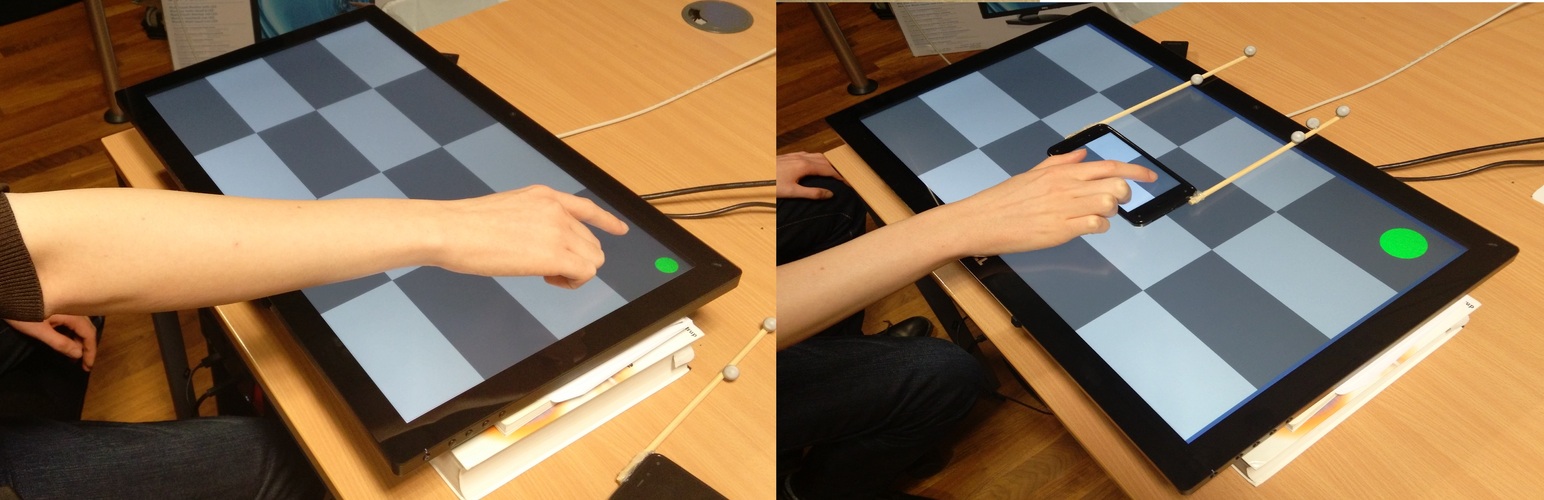}
    \caption{Two different C-D ratio cases of input.}
\end{figure}

The driving idea behind the work presented here is the investigation of various C-D ratio mappings in order to improve interaction over large distances using small input movements in mixed reality or VR. One critical component is the interaction with relatively small input devices (compared to the display space), as these imply a small motor space and small movements to traverse this space. Compare the two cases shown in figure 1, where the available motor space when using the smartphone totals less than 15cm on each coordinate axis, which can be utilized completely by finger, wrist or hand movements. In the case of the PC touch display, the user has to move the entire arm to reach corners of the display. On larger displays or VR applications that are not bound to physical screen size restrictions, the user might have to move with their feet to reach distant objects. This thesis explores the possibilities of controlling these large output spaces with small input movements, and to determine whether these smaller movements in motor space for small displays translates to faster or more accurate control than large movements. 

Touch-based input devices like smartphones have the limitation of a two-dimensional input space. Fingers can use this space, but can also reach several centimeters into the air. We seek to exploit this opportunity of expanding the motor space into the third dimension using optical tracking-based finger movement.

\section{Input methods} \label{sec:inputmethods}

Four modes of input are presented in the following. (1.) Basic large scale touch-screen input with C-D ratio of exactly one. Here, a PC monitor with touch input serves as a baseline test for situations where the user can directly select targets on the screen. (2.) Smartphone touch-screen input, used to explore different C-D ratio mappings with an input method that average users are likely to be familiar with. (3.) Z-Mapping: Finger-movement optical tracking-based input. Here, the user can move the environment around by moving the hand and finger. (4.) Z-Scaling: Environment zooming and navigation based on finger-movement. The user can zoom in and out of the environment by moving their finger up or down, scaling the distance needed to reach a target instead of modifying the C-D ratio. The following sections introduce more precise mathematical definitions of these input methods.

\subsection{Method 1: 1:1 mapped direct input} \label{method1}

Inputs such as physical touch locations can be directly mapped onto the virtual display position below the user's finger in a one to one fashion. It is defined as

\begin{equation}
f(\vec{p}) = \vec{p}, f: \mathbb{R}^2 \to \mathbb{R}^2 \label{eqdirect}
\end{equation} 

where \(\vec{p}\) is a two-dimensional input position. It serves as a baseline because it is functionally equivalent to having no C-D ratio mapping at all. 

\subsection{Method 2: 1:N mapping} \label{method2}

Method 1 can be extended such that the input position \(p_t\) is mapped to output depending on a custom C-D ratio mapping function \(H\). \(\vec{p}\) can be scaled by a factor \(N\) by introducing 

\begin{equation}
H(N) = \frac{1}{N}, H: \mathbb{R} \to \mathbb{R} \label{eq1nh}
\end{equation}

and modifying \(f\) in the following manner:

\begin{equation}
f(\vec{p}) = \vec{p} \cdot H(N), f: \mathbb{R}^2 \to \mathbb{R}^2
\end{equation}  

\subsection{Method 3: Z-Mapping} \label{method3}
\begin{figure*}
    \centering
    \includegraphics[width=1\linewidth]{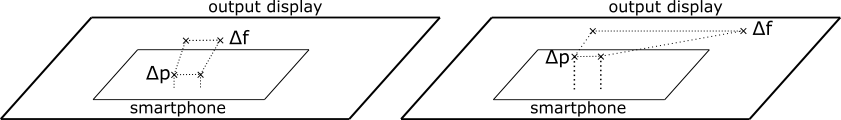}
    \caption[Z-Mapping]{ Fundamental principle of Z-Mapping. Left: A small input movement \(\Delta p \) a small distance above the surface results in a nearly identical output movement \(\Delta f \) on the target display. Right: The same input movement, but further up above the display results in a larger output movement }
    \label{fig:zmapping}
\end{figure*}

Z-Mapping extends the mappings above by extending the input space to three dimensions and using the extra coordinate to adjust the C-D ratio dynamically. The function \(H\) then operates on the new coordinate as input. The idea is to interpret the new dimension as input height, and to have users be able to smoothly modify the speed at which they control the environment by moving their fingers in arcs of different heights, higher arcs representing larger ''jumps'' in display space. Figure \ref{fig:zmapping} shows two examples. The final target selection still happens with touch input (typically on a smartphone). Compared to previous methods, \(f\) is extended with one parameter:

\begin{equation}
f(z, \Delta\vec{p} ) = \vec{f}_{t-1} + \Delta\vec{p}_{xy} \cdot H(p_z), f:\mathbb{R} \times \mathbb{R}^3 \to \mathbb{R}^2 
\end{equation}

where  \(\Delta\vec{p}\) is the change in input positions since the last frame, with \(\Delta\vec{p}_{xy}\) being the change in the x-y plane, \(\vec{f}_{t-1}\) is the mapped position of \(f\) from the previous frame, and \(H\) is the new C-D ratio mapping function that linearly modifies the 'C-D ratio based on movements in the new third dimension \(z\) of \(\vec{p}\):

\begin{equation}
H(z) = (1+\frac{z\cdot N}{M}), H: \mathbb{R} \to \mathbb{R}
\end{equation}

\(M\) and \(N\) are constants. The two functions are now described in detail: Since the resulting mapped position is not supposed to change when the input position is moved up or down, and is only supposed to change when the position is changed in the x and y directions, \(f\) formulates the output position relative to the previously mapped position \(\vec{f}_{t-1}\), meaning that if \(\Delta\vec{p}_{xy} = \vec{0}\) the result of the right term is \(\vec{0}\) and the output position does not change. \(H\) then only modifies the C-D ratio of input in the x-y directions. The goal of \(H\) is to modify the C-D ratio by a factor of \(M\)  for every \(N\) units which the user moves in the \(z\) direction. 

As an example from the context of a practical application, when a large one dimensional display space of size \(s_l\) is controlled with an input space that is limited to a smaller size \(s_s\), \(M\) and \(N\) might be chosen such that the maximum height a user's finger can reach above the smaller device changes the input ratio such that one movement across the entire range \(s_s\) maps onto a movement across the entire range of \(s_l\). In this case, N and M would be

\[N=h_{max} - h_{min}\]
\[M=\frac{s_l}{s_s}\]

where \(h_{max}, h_{min}\) are the maximum and minimum finger input height of the user, respectively, and the resulting function \(H\) is 

\begin{equation}
H(z) = (1+s_s \cdot  h_{max} -  h_{min}) \cdot \frac{z}{s_l}, H: \mathbb{R} \to \mathbb{R} \label{eq:zmapping}
\end{equation} This precise case is used in the user study experiment design for the evaluation of Z-Mapping. 

\subsection{Method 4: Z-Scaling}\label{method4}
\begin{figure*}
    \centering
    \includegraphics[width=0.8\linewidth]{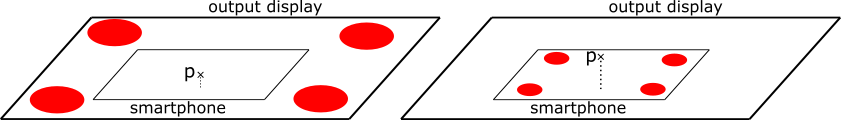}
    \caption[Z-Scaling example]{ Visualization of the scaling concept. Left: With the finger point \(\vec{p}\) at a low input height, the environment is visible near or at its original scale on the output display. Right: At a higher input height, the environment is scaled down, until it is eventually visible entirely on the smaller display. }
    \label{fig:zscaling}
\end{figure*}

The last examined input method does not change the C-D ratio directly, and instead relies on scaling the environment (zooming in and out) to reduce the amount of motor space that has to be traversed for some output movement. Figure \ref{fig:zscaling} shows a graphical representation of the finger scaling concept. The user first scales the environment down until a desired target is visible on the touch input screen, then traverses the necessary distance to the target in motor space, with the environment still scaled down, and then scales the environment back up, with the scaling happening relative to the target. After this, the target can be selected via touch input.

In this method, movements in motor space are mapped directly to movements in display space (similar to method \ref{method1}), with the driving idea being that reducing the amount of motor space that has to be covered to reach a certain target is functionally similar to increasing the C-D ratio such that the same motor space is covered but the output distance is increased instead, resulting in reduced movement times to reach targets.

The scaling used here is a simple uniform scale transformation, i.e. 

\[S = \begin{bmatrix}
s & 0 & 0 \\
0 & s & 0 \\
0 & 0 & s
\end{bmatrix}\]

where \(s\) is the scaling factor, with \(s < 1\) effectively ''zooming out'' and \(s > 1\) ''zooming in''. After applying this transformation, any point \(\vec{p}\) in the environment is shifted to the new location 

\begin{equation}
\vec{p_s} = S \cdot (\vec{p} - \vec{c}) + \vec{c}  = s \cdot (\vec{p} - \vec{c}) + \vec{c} \label{eqscaleshift}
\end{equation} 

where \( \vec{c} \) is the center position of the environment. 

In order to specify a target location, the foot \(\vec{p_f}\) of the perpendicular from the users finger position \(\vec{p_i}\) to the touch screens plane (typically from a smartphone), which is centered at \(\vec{p_s}\) with plane normal vector \(\vec{n}\), is used:

\begin{equation}
\vec{p_f} = \vec{p_i} - ((\vec{p_i} - \vec{p_s}) \bullet \vec{n}) \cdot \vec{n}
\label{eqfoot} \end{equation}

The equation above computes the foot by subtracting the part of the input user's position vector that is perpendicular to the touch screen display's plane normal vector. From the user's perspective, the target is on the smartphone screen location ''below the finger''. The scaling is then applied such that the location at the foot of this perpendicular remains the same after scaling, i.e. the user's target remains below the finger. This is achieved by translating the entire environment in the opposite direction corresponding to how much the target location moved after scaling, i.e. the environment's center \(\vec{c}\) is moved to its post-scale location \(\vec{c_s}\) via

\begin{equation} \vec{c_s} = \vec{c} - (\vec{p_{fs}} - \vec{p_f})
\label{move} \end{equation}

where \(\vec{p_f}\) is the perpendicular's foot location before scaling, and \(\vec{p_{fs}}\) is its location after scaling which can be determined using equations \ref{eqscaleshift} and \ref{eqfoot} above. In the default implementation of z-scaling the above is applied to both upwards and downwards motions of the user's finger, i.e. the target location is kept under the finger when zooming in as well as when zooming out, which allows the user to finely adjust their target positioning by moving their fingers up and down in small intervals, for example to improve landing accuracy on the target. 

\section{Practical Applications}\label{sec:applications}
\begin{figure}
    \centering
    \includegraphics[width=0.8\linewidth]{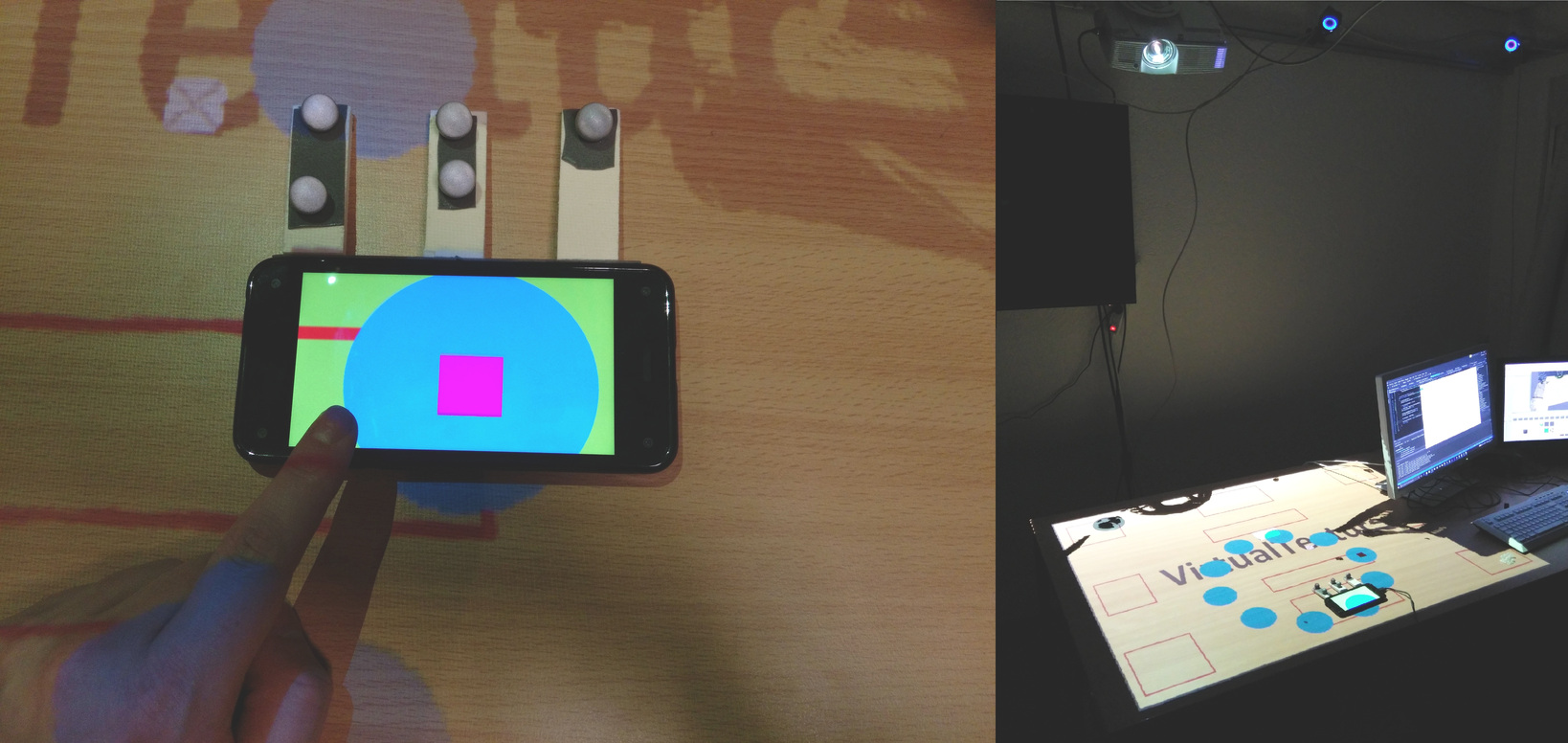}
    \caption[Spatial augmented reality environment example]{SAR environment displaying a basic prototyping environment. }
    \label{fig:spatialsetup}
\end{figure}

\begin{figure}
    \centering
    \includegraphics[width=0.7\linewidth]{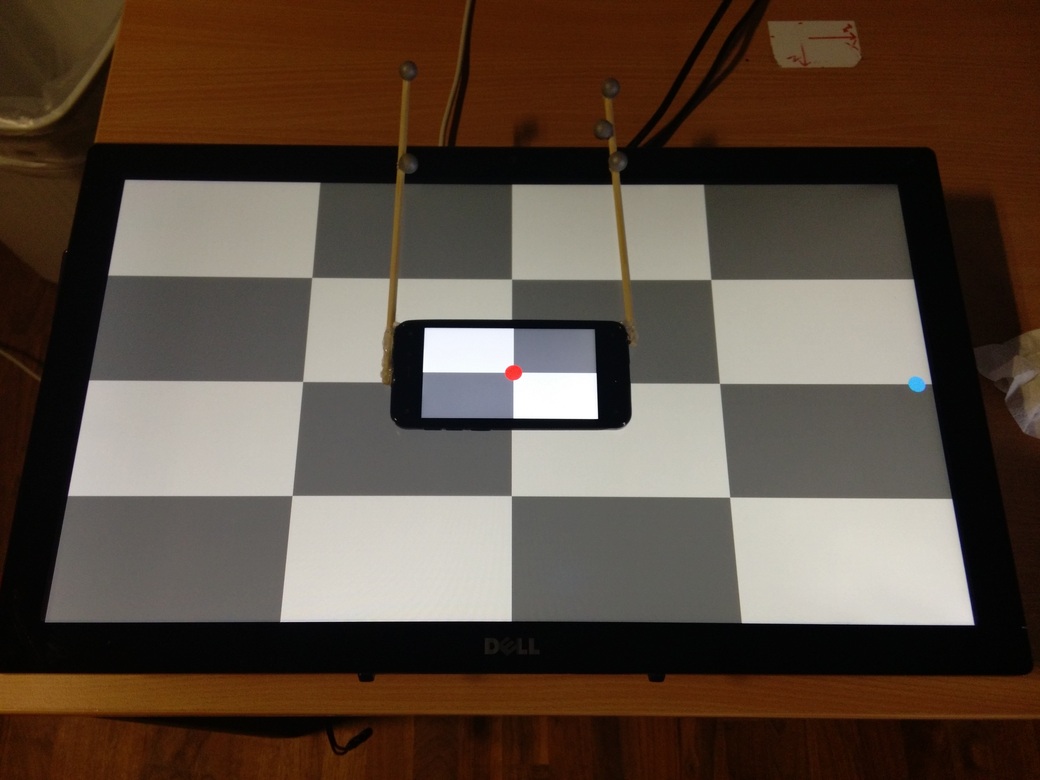}
    \caption[User study environment]{Running example of the multi-display environment of the user study.  }
    \label{fig:userstudysetup}
\end{figure}

The described input methods are implemented in two practical applications, an SAR environment seen in figure \ref{fig:spatialsetup}, which was used for prototyping and development, and a separate setup used in the user study seen in figure \ref{fig:userstudysetup}, where a PC touch monitor is used for input and as the larger output space instead of a projector. Both setups share other hardware and configuration aspects, such as a central smartphone that displays a cutout of the larger environment. Targets are shown and selected via touch input.

\section{Implementation}\label{sec:implementation}
The following sections will details about the various components involved in the implementation of the input methods described previously.

\subsection{Hardware} \label{sec:hardware}

Deployed hardware components consist of the following parts: 

\begin{itemize}
    
    \item  Microsoft Kinect (v2) which is mainly used during calibration by the RoomAlive toolkit to recreate a virtual model of the real-life working space, most prominently featuring the desk upon which the experiments are conducted. The Kinect is only used for 3D scanning, and not used during the actual runtime of the Unity applications, as optical tracking is used to determine locations of real world objects such as fingers and smartphones.
    \item  Optoma GT 1070 projector which is used both in the calibration process of the RoomAlive toolkit, as well to project the final running application environment onto the real world workspace. 
    \item  Eight OptiTrack Prime 13 camera sensor units used for optical marker tracking.
    \item 22 inch Dell S2340T PC touch screen, required for the user study.
    \item  Workstation consisting of a wooden desk that is used as a display surface for the projector used in the user study and input method prototyping, and a consumer PC used for developing input methods as well as running the actual user study Unity applications. 
    \item  PC hardware which consists of an Intel Xeon E5-2620 CPU with 6 cores of 2.00 GHz base frequency each, an NVIDIA GTX 1070 graphics card, and 32GB of DDR3 RAM. 
\end{itemize}

\subsection{Optitrack object and finger tracking}

Optitrack and its control software allows the definition of so-called rigid bodies, a user-defined collection of markers in the real world whose position and orientation relative to each other remains fixed (i.e. they cannot be deformed). In addition to allowing the tracking of rotations which is not possible with a single marker due to lack of available degrees of freedom, tracking a single object with multiple markers also increases reliability of tracking as a rigid body's location can be tracked even when one or multiple of its markers are not visible to the cameras, although a minimum count of 3 markers is generally still required. In the following, the exact marker setup for the tracked smartphone as well as for finger tracking are described.

\subsubsection{Smartphone Rigid Body} \label{smartphonerigidbody}
\begin{figure}
    \centering
    \includegraphics[width=0.7\linewidth]{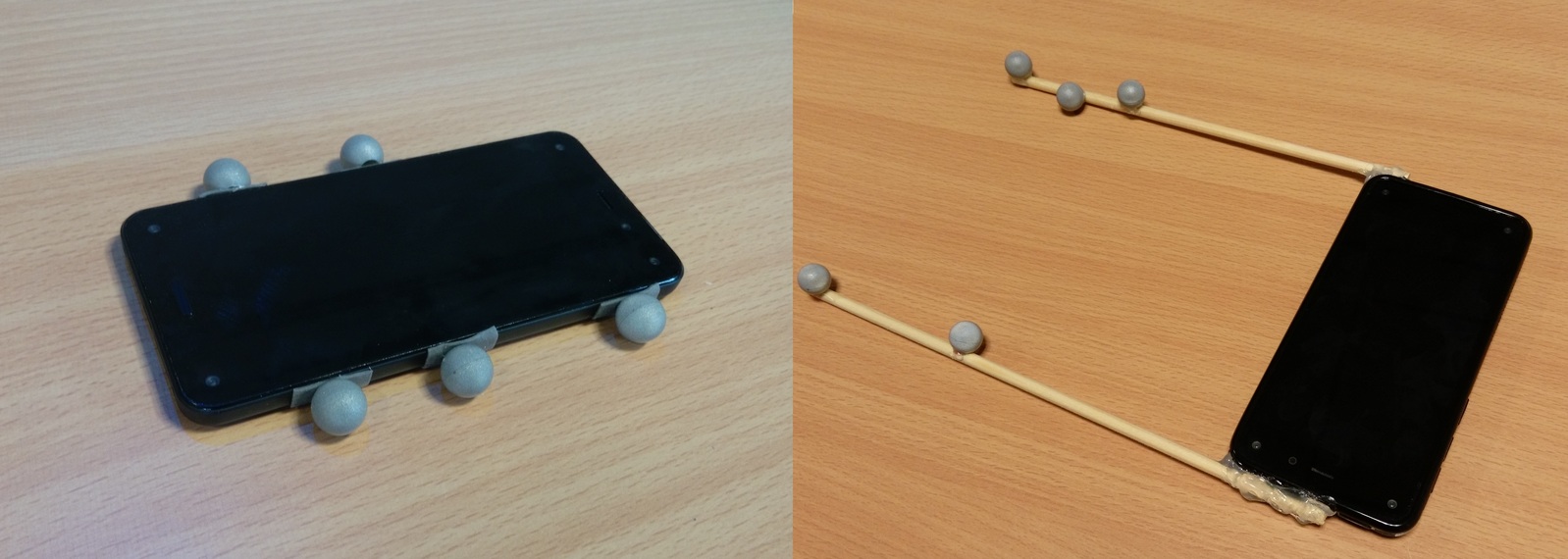}
    \caption[Smartphone rigid body comparisons]{Two rigid body configurations side-by-side.}
    \label{fig:smartphonerigidbody}
\end{figure}

A rigid body is defined in Optitrack to allow tracking of the smartphone. Two possible solutions are shown in figure \ref{fig:smartphonerigidbody} side by side, with the setup on the right being used in the user study. The setups are tradeoffs: The left configuration can cause problems when a user's hand occludes markers partially or completely, and the setup on the right is mechanically more bulky, which is undesirable for mobile setups (which is not a problem in the user study where all hardware components are static).

\subsubsection{Finger tracking} \label{sec:fingertracking}
\begin{figure}
    \centering
    \includegraphics[width=0.3\linewidth]{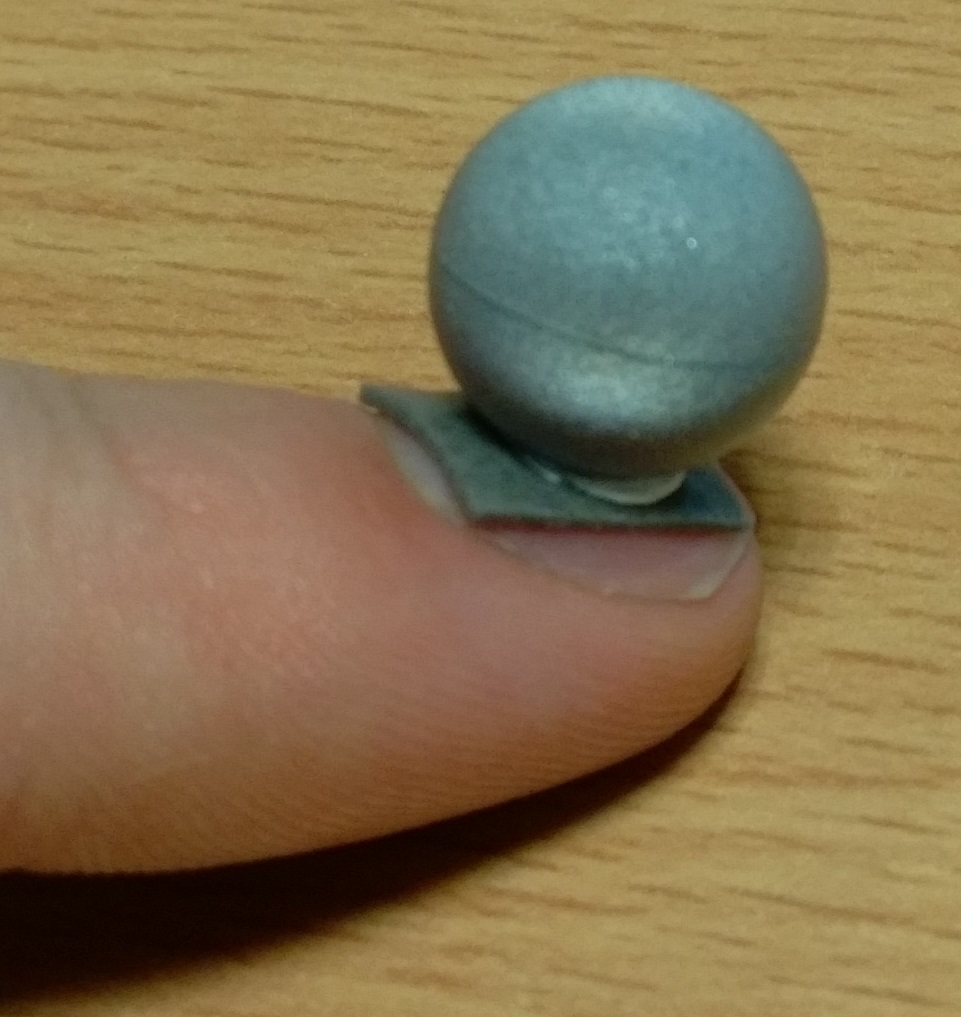}
    \caption[Finger marker]{Marker attached to finger via double sided bonding tape.}
    \label{fig:fingermarker}
\end{figure}
One finger of the user is tracked so it can be used with input methods from sections \ref{method3} and \ref{method4}. A single spherical marker is used (see figure \ref{fig:fingermarker}) and attached to the finger via double sided bonding tape. The set of markers read from the Optitrack system are filtered according to application-defined logic to identify the relevant finger marker when other markers are visible.

\subsection{Transformation acquisition}
\begin{figure}
    \centering
    \includegraphics[width=0.4\linewidth]{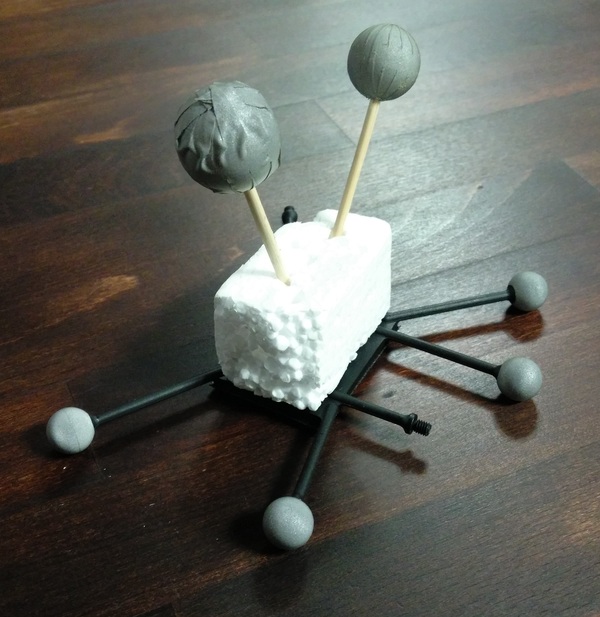}
    \caption["Calibrator" rigid body]{ Special purpose rigid body that is used to represent a coordinate system transform from the Optitrack coordinate system to the RoomAlive Toolkit's and Unity's coordinate system. }
    \label{fig:calibrator}
\end{figure}

In order to figure out the transformation between the Optitrack and Unity coordinate system, a custom calibrator rigid body (see \ref{fig:calibrator}) is aligned with the Kinects position in the real world (which serves as Unity's coordinate origin) and its orientation\(Q_{T}\) and position \(\vec{x}_{T}\) are used as the rotation and translation part of the coordinate transform.

\subsubsection{Transformation usage}
After acquiring rotation quaternion \(Q_{T}\) and translation vector \(\vec{x}_{T}\), they are applied by

\[Q' = Q_T^{-1} \cdot Q \]

where\(Q\) and  \(Q'\) are an objects rotation before and after the transformation, respectively. The final position \(\vec{x}'\) with original position \(\vec{x}\) is given by

\[\vec{x}' = Q_T^{-1} \cdot ( \vec{x} - \vec{x}_{T} ) + \vec{x}_{T} \]

\section{Evaluation}\label{sec:evaluation}
Several C-D ratio mappings presented previously were subjected to a user study in order to compare their performance. Three candidates (direct 1:1 mapping via smartphone touch dragging, Z-Mapping and Z-Scaling) were initially selected based on their performance during their development phase. In addition, the 1:1 mapped direct PC touch monitor input method (see figure \ref{fig:twocases}) was added as a baseline. Z-Mapping and Z-Scaling were then compared in the user study environment as part of an informal pilot study with two expert users. As a result of that pilot study, Z-Mapping was chosen as the novel input method to be tested with more subjects, due to higher performance of both users in the pilot study. The following section describes the conducted experiment, presents results and then offers interpretation of these results. 

\subsection{Experiment description}
As an evaluation environment, the setup described in section \ref{sec:applications} was used. Subjects were asked to acquire targets using three different input methods: 

\begin{itemize}
    \item Direct (1:1 C-D ratio mapped) PC-display touch input, where users could physically select targets on the screen. 
    \item 1:1 C-D ratio mapped smartphone touch-input, where users had to move targets into the view of the smartphone screen by dragging environment towards it using touch input.
    \item Z-Mapping as described in section \ref{method3}, where users had to move targets into view of the smartphone screen by dragging the environment towards it using finger movements in the air. The minimum and maximum height as described in section \ref{method3} were acquired using a simple height calibration where users placed their fingers at the lowest and highest possible position they could reach with their fingers with a hand placed down on the large PC touch screen, which were then recorded.
\end{itemize}

\subsection{Procedure}
After initial introductions and signing of a participant agreement waiver and release form, subjects filled out a demographics questionnaire. All subjects had a marker attached to their index finger that was used for optical tracking as described in section \ref{sec:fingertracking}. After explaining the basic experiment setup and rules, for each input method, subjects then executed a target acquisition task (described below), where circular targets had to be selected on one or two of the used displays. Before each input method, subjects were provided with a short demonstration of the respective input technique's concept. Subjects then executed an initial training sequence in order to familiarize themselves with the input techniques, followed by a main run as described below in section \ref{sec:experimentdesign}. Subjects executed the task in a seated position in front of the PC touch screen. After each input method, participants were asked to fill out an after-scenario-questionnaire \cite{lewis1991psychometric} and the NASA TLX questionnaire \cite{hart1988development}, and a preference questionnaire after all input method tasks were completed. Finally, an informal interview was held in order to allow each subject to expand on their opinion of conducted tasks in their own words.

\subsubsection{Target acquisition task}
Circular targets are displayed on the two used displays (PC touch screen, and smartphone screen).  One display was used for the PC-display touch input case, and two displays for both other input methods, where the smartphone is placed centrally on the larger screen and displays a cutout of its environment. Initially, both a red and blue target are displayed. Upon clicking on the red target, the blue target is activated for selection and turned green, after which users can navigate to it and acquire it using direct touch input (for all methods). In the smartphone touch input case and Z-Mapping, users first have to relocate the target such that it is visible on the smartphone display, as PC display touch-screen input was disabled for those input methods. After acquiring a green target, the environment is reset to its starting location and the red target, in addition to the next blue target, is displayed again. The purpose of this starting position is for the user to be able to take time and visually recognize the position of the new target, as movement times and accuracy are not tracked during this period. This process repeats until all required samples have been gathered. Subjects were specifically instructed to first locate the new (blue) target, and only after visually acquiring it, select the red target and then select the green target. This is to prevent targer search times from entering into the recorded movement times.

\subsubsection{Design} \label{sec:experimentdesign}
The experiment used a repeated measures within-subject design. The independent variables were input method type(direct-touch-PC, direct-touch-smartphone, z-mapping), the direction in which targets appear (8 directions in total, each 45 degrees apart in a circle around the large display center), as well as the index of difficulty of those targets resulting from their sizes and distances to the smartphone (average of 15 targets for each direction). 120 targets had to be selected per input method with 360 main samples gathered over 3 input methods in total for each subject, in addition to 180 training samples for each subject over all three input methods. For the index of difficulty, values of 2-5 were targeted, but with a minimum target width of 1cm, the highest possible \(ID\) values reachable were 4.85 (in the corners of the PC touch screen display) and 4.67 (near the edge of the PC touch screen on the right and left sides in the middle), which were used to represent the \(ID\) category 5.
Participants performed the procedure described above in one session lasting roughly 50 minutes. Each of the three input methods was performed one after the other, with counterbalanced order permutations. The main sequence of 120 samples per input method was executed in 4 blocks of 30 targets each. Each subject selected the same targets, but the order of targets was randomized per-subject using selection without replacement. After each block in both the training and main sequences, the subjects were offered an optional short break in case of fatigue. 

\subsubsection{Participants}
A total of 20 (4 female, 16 male) subjects participated in the study, in the age range of 21-34. 14 subjects worked in the field of or studied computer science,  2 subjects were electrical engineers, 3 subjects worked or studied in health sciences and 1 subject was a product designer. 9 subjects were visually impaired, with all of them using corrective glasses or contact lenses. 18 subjects were right-handed, and 2 subjects left handed. All 20 subjects specified that they use smartphones on a daily basis. Subjects were volunteers, but with the exception of one were awarded with a gift certificate worth 10 euros (the subjects were told of this when signing up for the test).

\subsubsection{Apparatus}
The study was executed on a PC running the target acquisition test software with an Intel Xeon E5-2620 CPU with 6 cores of 2.00 GHz base frequency each, an NVIDIA GTX 1070 graphics card, and 32GB of DDR3 RAM. An Amazon Fire smartphone was used as the smartphone device, and a 2 inch Dell S2340T PC touch screen display served as the larger display and for the PC touch screen input method. The Optitrack M13 based optical tracking environment is described in section \ref{sec:hardware}.

\subsection{Results}\label{sec:results}
The following section presents results of the experiment. For the indices of difficulty range of 2 through 5, four \(ID\) categories are defined. They are: Category 2 with \(1.5 < ID \leq 2.5\), category 3 with \(2.5 < ID \leq 3.5\), category 4 with \(3.5 < ID \leq 4.5\) and category 5 with \(4.5 < ID \leq 5.5\). Data was also examined by each executed 30 sample block. User preferences as determined by the preference questionnaire are also presented. The ranks given by subjects in their answers are converted to ordinal samples using their ranking number (from 1 = best, to 3 = worst). These results are then tested for statistical significance.

\subsubsection{Statistical significance testing} \label{sec:significance}
Samples from each grouping (overall, by ID, by block) were first averaged per user and then subjected to significance testing, resulting in sample count \(n=20\) for each input method. Normality was assumed and tested using the Shapiro-Wilkes normality test \cite{shapiro1965analysis} with \(\alpha = 0.05\). ANOVA was used to determine statistical significance for normally distributed samples, and Friedman test \cite{friedman1937use} was used for non-normally distributed samples, both with \(\alpha = 0.05\). When passed, the alpha level was adjusted for multiple comparisons using conservative Bonferroni correction \cite{dunn1961multiple}, with resulting \(\alpha_b = 0.017\). Finally, post-hoc tests for normal data used paired Student's t-tests \cite{student1908probable}, and Wilcoxon signed-rank tests \cite{wilcoxon1945individual} for non normal data, both with the modified alpha level resulting from Bonferroni correction. 

The tested hypothesis was that there is a relationship (in movement time or accuracy, respectively) between on the input method used to perform the task and the resulting metric being measured. In the case of the preference questionnaire, the hypothesis tested for is that users have differing preferences between input methods with respect to certain criteria (perceived speed and accuracy, as well as overall preference). 

\subsubsection{Data Collection and Analysis}
Two primary types of data are collected and used to judge subject performance: Movement time and accuracy. Movement time is recorded from touch-up of the red target to touch-up of the green target.The accuracy of task execution \(a\) is defined by the number of a subject's average hits \(n_h\) and misses \(n_m\):

\[a = \frac{n_h}{n_h + n_m}\]

\subsubsection{Accuracy and movement times}

\begin{figure}
    \centering
    \includegraphics[width=0.7\linewidth]{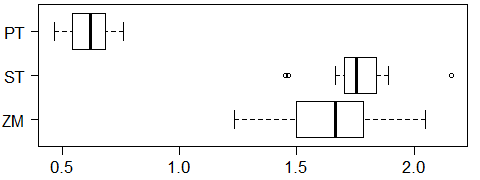}
    \caption[Movement time box plot over all gathered samples. ]{Box plot showing summarized movement times of different input methods over all blocks and indices of difficulty.}
    \label{fig:mtavgboxplot}
\end{figure}

\begin{figure}
    \centering
    \includegraphics[width=1.0\linewidth]{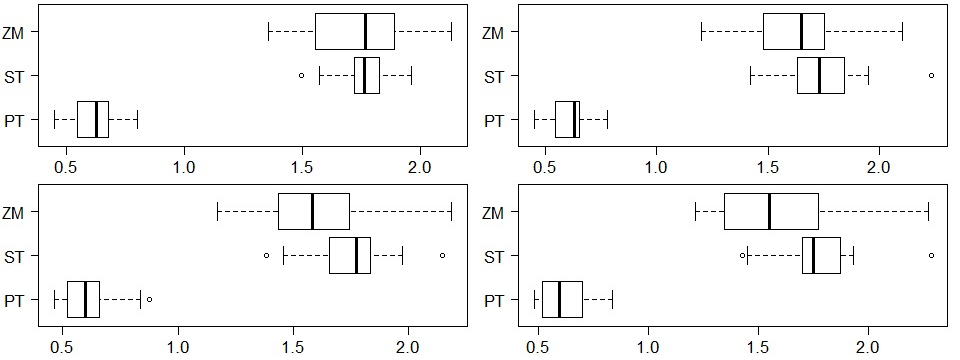}
    \caption[Movement time box plots, for blocks 1 through 4. ]{Box plots showing movement times of different input methods for separated blocks 1 (top left) through 4 (bottom right), proceeding left to right, top to bottom.}
    \label{fig:mtblocksboxplot}
\end{figure}

\begin{figure}
    \centering
    \includegraphics[width=1.0\linewidth]{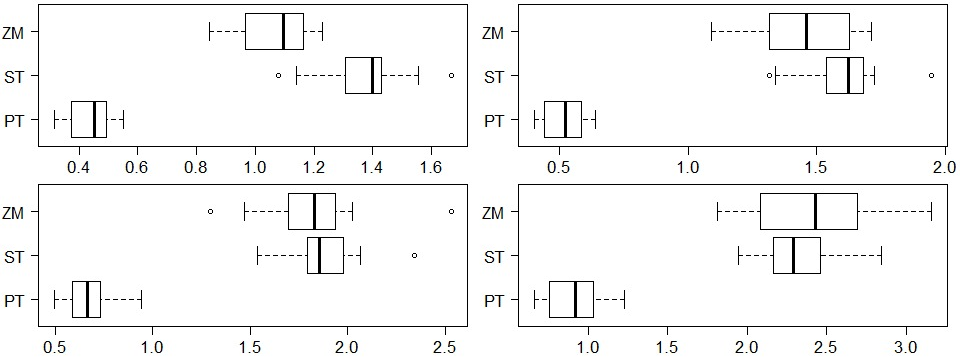}
    \caption[Movement time box plots, for ID categories 2 through 5. ]{Box plots showing movement times of different input methods for separated ID categories 1 (top left) through 4 (bottom right), proceeding left to right, top to bottom.}
    \label{fig:mtidsboxplot}
\end{figure}

\begin{figure}
    \centering
    \includegraphics[width=0.7\linewidth]{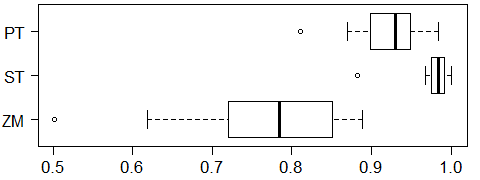}
    \caption[Accuracy box plot over all gathered samples. ]{Box plot showing summarized accuracies of different input methods over all blocks and indices of difficulty.}
    \label{fig:mtavgboxplotacc}
\end{figure}

\begin{figure}
    \centering
    \includegraphics[width=1.0\linewidth]{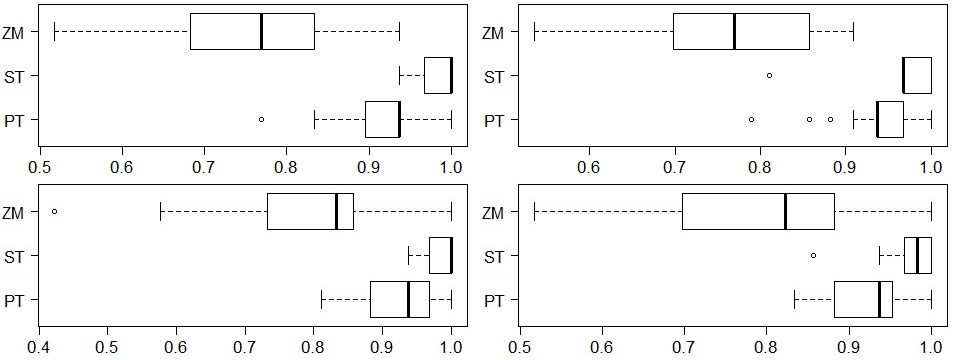}
    \caption[Accuracy box plots, for blocks 1 through 4. ]{Box plots showing accuracy of different input methods for separated blocks 1 (top left) through 4 (bottom right), proceeding left to right, top to bottom.}
    \label{fig:mtblocksboxplotacc}
\end{figure}

\begin{figure}
    \centering
    \includegraphics[width=1.0\linewidth]{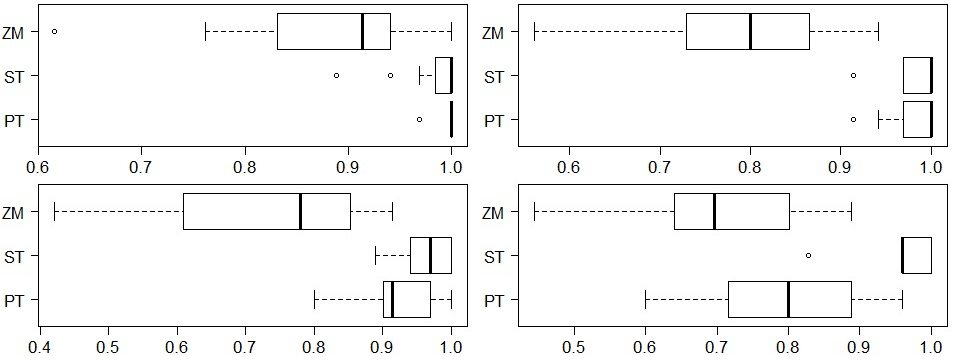}
    \caption[Accuracy box plots, for ID categories 2 through 5. ]{Box plots showing accuracy of different input methods for separated ID categories 1 (top left) through 4 (bottom right), proceeding left to right, top to bottom.}
    \label{fig:mtidsboxplotacc}
\end{figure}

Figures \ref{fig:mtavgboxplot} through \ref{fig:mtidsboxplotacc} show results of average means and standard deviations for movement times and accuracy. Tables \ref{tab:mtnormality} through \ref{tab:accposthoc} show results of significance tests for both metrics. Input methods are abbreviated: ZM = Z-Mapping, PT = PC touch input, ST = smartphone touch input. 

\begin{table*}
    \begin{center}
        \begin{tabular}{ | c | c | c | c | }
            \hline
            & ZM & ST & PC \\ \hline
            Overall & \(\mu_{zm} = 1.64s, \sigma_{zm} = 0.20 \) & \(\mu_{st} = 1.76s, \sigma_{st} = 0.15\) &  \(\mu_{pt}= 0.62s, \sigma_{pt} = 0.09\)\\ \hline
        \end{tabular}
        \caption[Means and standard deviations for overall movement time data]{Means and standard deviations for movement times for overall data. } \label{tab:mtoverall}
    \end{center}
\end{table*}

 \begin{table*}
    \begin{center}
        \begin{tabular}{ | c | c | c | c | }
            \hline
            & ZM & ST & PC \\ \hline
            Block 1 & \(\mu = 1.74, \sigma = 0.24\) & \(\mu = 1.76, \sigma = 0.11\) &  \(\mu = 0.62, \sigma = 0.10\)\\ \hline
            Block 2 &  \(\mu = 1.62, \sigma = 0.21\) &  \(\mu = 1.74, \sigma = 0.19\) &  \(\mu = 0.61, \sigma = 0.09\)\\ \hline
            Block 3 &  \(\mu = 1.60, \sigma = 0.24\) &  \(\mu = 1.75, \sigma = 0.18\) &  \(\mu = 0.61, \sigma = 0.12\) \\ \hline
            Block 4 &  \(\mu = 1.60, \sigma = 0.30\) &  \(\mu = 1.77, \sigma = 0.18\) &  \(\mu = 0.62, \sigma = 0.11\)\\ \hline
        \end{tabular}
        \caption[Means and standard deviations for movement times by block]{Means and standard deviations for movement times by block. } \label{tab:mtblock}
    \end{center}
\end{table*}

\begin{table*}
    \begin{center}
        \begin{tabular}{ | c | c | c | c | }
            \hline
            & ZM & ST & PC \\ \hline
            ID category 2 &  \(\mu = 1.07, \sigma = 0.12\) &  \(\mu = 1.37, \sigma = 0.13\) &  \(\mu = 0.43, \sigma = 0.07\)\\ \hline
            ID category 3 &  \(\mu = 1.46, \sigma = 0.19\) &  \(\mu = 1.61, \sigma = 0.14\) &  \(\mu = 0.52, \sigma = 0.08\)\\ \hline
            ID category 4 &  \(\mu = 1.82, \sigma = 0.25\) &  \(\mu = 1.88, \sigma = 0.18\) &  \(\mu = 0.66, \sigma = 0.11\) \\ \hline
            ID category 5 &  \(\mu = 1.24, \sigma = 0.39\) &  \(\mu = 2.31, \sigma = 0.22\) &  \(\mu = 0.92, \sigma = 0.16\)\\ \hline
        \end{tabular}
        \caption[Means and standard deviations for movement times by ID categories]{Means and standard deviations for movement times by ID categories. }\label{tab:mtid}
    \end{center} 
\end{table*}

 \begin{table*}
    \begin{center}
        \begin{tabular}{ | c | c | c | c | }
            \hline
            & ZM & ST & PC \\ \hline
            Overall & \(\mu_{zm} = 0.77, \sigma_{zm} = 0.10 \) & \(\mu_{st} = 0.98, \sigma_{st} = 0.02\) & \(\mu_{pt}= 0.92, \sigma_{pt} = 0.04\)\\ \hline
        \end{tabular}
        \caption[Means and standard deviations for overall accuracy data]{Means and standard deviations for accuracy for overall data. } \label{tab:mtoverallacc}
    \end{center}
\end{table*}

\begin{table*}
    \begin{center}
        \begin{tabular}{ | c | c | c | c | }
            \hline
            & ZM & ST & PC \\ \hline
            Block 1 &  \(\mu = 0.76 , \sigma = 0.10\) &  \(\mu = 0.98, \sigma = 0.03\) &  \(\mu = 0.92, \sigma = 0.05\)\\ \hline
            Block 2 &  \(\mu = 0.77, \sigma = 0.11\) &  \(\mu = 0.97, \sigma = 0.04\) &  \(\mu = 0.94, \sigma = 0.05\)\\ \hline
            Block 3 &  \(\mu = 0.79, \sigma = 0.13\) &  \(\mu = 0.98, \sigma = 0.02\) &  \(\mu = 0.93, \sigma = 0.06\) \\ \hline
            Block 4 &  \(\mu = 0.79, \sigma = 0.13\) &  \(\mu = 0.98, \sigma = 0.03\) &  \(\mu = 0.92, \sigma = 0.05\)\\ \hline
        \end{tabular}
        \caption[Means and standard deviations for accuracy by block]{Means and standard deviations for accuracy by block. } \label{tab:mtblockacc}
    \end{center}
\end{table*}

\begin{table*}
    \begin{center}
        \begin{tabular}{ | c | c | c | c | }
            \hline
            & ZM & ST & PC \\ \hline
            ID category 2 &  \(\mu = 0.89, \sigma = 0.09\) &  \(\mu = 0.99, \sigma = 0.03\) &  \(\mu = 0.99, \sigma = 0.01\)\\ \hline
            ID category 3 &  \(\mu = 0.79, \sigma = 0.10\) &  \(\mu = 0.99, \sigma = 0.02\) &  \(\mu = 0.98, \sigma = 0.03\)\\ \hline
            ID category 4 &  \(\mu = 0.74, \sigma = 0.14\) &  \(\mu = 0.97, \sigma = 0.03\) &  \(\mu = 0.92, \sigma = 0.05\) \\ \hline
            ID category 5 &  \(\mu = 0.70, \sigma = 0.13\) &  \(\mu = 0.97, \sigma = 0.04\) &  \(\mu = 0.80, \sigma = 0.01\)\\ \hline
        \end{tabular}
        \caption[Means and standard deviations for accuracy by ID categories]{Means and standard deviations for accuracy by ID categories. }\label{tab:mtidacc}
    \end{center} 
\end{table*}

\begin{table*}
    \begin{center}
        \begin{tabular}{ | c | c | c | c | c | c | c | c | c | c | }
            \hline
            & Overall	& Block 1 	& Block 2 	& Block 3	& Block 4 & ID 2 & ID 3 & ID 4 	& ID 5 \\ \hline
            ZM & 0.9917 	&  0.40 	& 0.91 	&  0.84	 &  0.14 & 0.09 & 0.28 	& 0.08 	& 0.51 \\ \hline
            ST & 0.06053 	&  0.63	    & 0.45	&  0.38  &  0.14 & 0.28 & 0.18 	& 0.43 	& 0.78 \\ \hline
            PT & 0.3105 	&  0.36 	& 0.16	&  0.15  &  0.18 & 0.18	& 0.09 	& 0.61 	& 0.42  \\ \hline
        \end{tabular}
        \caption[Results of normality testing using Shapiro-Wilk test on movement time groups]{p-values for Shapiro-Wilk tests of each input method and sample group in order to test for normality. } \label{tab:mtnormality}
    \end{center}
\end{table*}

\begin{table}
    \begin{center}
        \begin{tabular}{ | c | c | c | }
            \hline
            & ANOVA p-value & Passed? \\ \hline
            Overall 		& \( < 2.2 \cdot 10^{-16}\)  &  \checkmark   \\ \hline
            Block 1 		& \( < 2.2 \cdot 10^{-16}\)  &  \checkmark   \\ \hline
            Block 2 		& \( 4.19 \cdot 10^{-14}\)  &  \checkmark   \\ \hline
            Block 3 		& \( 2.58\cdot 10^{-13}\)  &  \checkmark     \\ \hline
            Block 4  		& \( 1.36 \cdot 10^{-12}\)  &  \checkmark   \\ \hline
            ID category 2  	& \( < 2.2 \cdot 10^{-16}\)  &  \checkmark   \\ \hline
            ID category 3   & \( < 2.2 \cdot 10^{-16}\)  &  \checkmark    \\ \hline
            ID category 4  	& \( 1.48 \cdot 10^{-15}\)  &  \checkmark     \\ \hline
            ID category 5   & \( 9.25 \cdot 10^{-11}\)  &  \checkmark  \\ \hline
        \end{tabular}
        \caption[ANOVA test results for movement times. ]{ANOVA test results for movement times. All groupings pass the test.} \label{tab:mtsignificance}
    \end{center}
\end{table}

\begin{table*}
    \begin{center}
        \begin{tabular}{ | c | c | c | c | c |}
            \hline
            Group & Method 1 & Method 2 & t-test p-value & Passed? \\ \hline
            \multirow{3}{*}{Overall} 	& ZM & ST & \( 0.05 \)  &  x   \\ \cline{2-5}
            & ZM & PT &  \( 3.78 \cdot 10^{-14}\) & \checkmark \\ \cline{2-5}
            & PT & ST &  \( < 2.20 \cdot 10^{-16}\) & \checkmark \\ \hline
            \multirow{3}{*}{Block 1} 	& ZM & ST & \( 0.69 \)  &  x   \\ \cline{2-5}
            & ZM & PT &  \( 3.26 \cdot 10^{-13}\) & \checkmark \\ \cline{2-5}
            & PT & ST &  \( < 2.20 \cdot 10^{-16}\) & \checkmark \\ \hline
            \multirow{3}{*}{Block 2} 	& ZM & ST & \( 0.03 \)  &  x   \\ \cline{2-5}
            & ZM & PT &  \( 1.50 \cdot 10^{-13}\) & \checkmark \\ \cline{2-5}
            & PT & ST &  \( 2.61 \cdot 10^{-16}\) & \checkmark \\ \hline
            \multirow{3}{*}{Block 3} 	& ZM & ST & \( 0.05 \)  &  x   \\ \cline{2-5}
            & ZM & PT &  \( 2.60 \cdot 10^{-12}\) & \checkmark \\ \cline{2-5}
            & PT & ST &  \( 6.00 \cdot 10^{-16}\) & \checkmark \\ \hline
            \multirow{3}{*}{Block 4} 	& ZM & ST & \( 0.03\)  &  x   \\ \cline{2-5}
            & ZM & PT &  \( 1.07 \cdot 10^{-11}\) & \checkmark \\ \cline{2-5}
            & PT & ST &  \( < 2.20 \cdot 10^{-16}\) & \checkmark \\ \hline
            \multirow{3}{*}{ID Category 2} 	& ZM & ST & \( 3.14 \cdot 10^{-8}\)  &  \checkmark   \\ \cline{2-5}
            & ZM & PT &  \( 2.13 \cdot 10^{-14}\) & \checkmark \\ \cline{2-5}
            & PT & ST &  \( <2.20 \cdot 10^{-16}\) & \checkmark \\ \hline
            \multirow{3}{*}{ID Category 3} 	& ZM & ST & \( 0.01 \)  &  \checkmark   \\  \cline{2-5}
            & ZM & PT &  \( 3.283 \cdot 10^{-14}\) & \checkmark \\ \cline{2-5}
            & PT & ST &  \( < 2.20 \cdot 10^{-16}\) & \checkmark \\ \hline
            \multirow{3}{*}{ID Category 4} 	& ZM & ST & \( 0.46 \)  &  x   \\ \cline{2-5} 
            & ZM & PT &  \(  1.296 \cdot 10^{-13}\) & \checkmark \\ \cline{2-5}
            & PT & ST &  \( < 2.20 \cdot 10^{-16}\) & \checkmark \\ \hline
            \multirow{3}{*}{ID Category 5} 	& ZM & ST & \( 0.37 \)  &  x   \\ \cline{2-5}
            & ZM & PT &  \( 8.943 \cdot 10^{-12}\) & \checkmark \\ \cline{2-5}
            & PT & ST &  \( 1.741 \cdot 10^{-15}\) & \checkmark \\ 
            \hline
        \end{tabular}
        \caption[Results of paired t-tests between all combinations of input methods for movement times. ]{Results of paired t-tests between all combinations of input methods for movement times. } \label{tab:mtposthoc}
    \end{center}
\end{table*}

\begin{table}
    \footnotesize
    \begin{center}
        
        \begin{tabular}{ | c | c | c | c | }
            
            \hline
            & ZM	& ST 	& PT \\ \hline
            Overall & \( 0.09\) &  \( 8.83 \cdot 10^{-7}\) & \( 0.22 \)	\\ \hline
            Block 1 & \( 0.95\) &  \( 4.04 \cdot 10^{-5}\) & \( 0.01 \)\\ \hline
            Block 2 & \( 0.02\) &  \( 5.18 \cdot 10^{-7}\) & \( 4,00*  \cdot 10^{-3}\)  \\ \hline
            Block 3 & \( 0.09\) &  \( 7.57 \cdot 10^{-5}\) & \( 0.05 \)  \\ \hline
            Block 4 & \( 0.76\) &  \( 1.18 \cdot 10^{-5}\) & \( 0.14 \)  \\ \hline
            ID Category 2 & \( 0.01\) &  \( 7.91 \cdot 10^{-7}\) & \(  2.89 \cdot 10^{-7} \) \\ \hline
            ID Category 3 & \( 0.29\) &  \(6.55 \cdot 10^{-6}\) & \( 8.91 \cdot 10^{-5} \) \\ \hline
            ID Category 4 & \( 0.07\) &  \( 0.001 \) & \( 0.16 \)     \\ \hline
            ID Category 5 & \( 0.45\) &  \(  3.76 \cdot 10^{-6}\) & \( 0.32 \) \\ \hline
            
        \end{tabular}
        \caption[Results of normality testing using Shapiro-Wilk test on accuracy groups]{p-values for Shapiro-Wilk tests of each input method. } \label{tab:accnormality}
        
    \end{center}
\end{table} 

\begin{table}
    \begin{center}
        \begin{tabular}{ | c | c | c | }
            \hline
            & Friedman p-value & Passed? \\ \hline
            Overall 		& \( 5.33 \cdot 10^{-9}\)  &  \checkmark   \\ \hline
            Block 1 		& \( 2.40 \cdot 10^{-8}\)  &  \checkmark   \\ \hline
            Block 2 		& \(  4.03 \cdot 10^{-8}\)  &  \checkmark   \\ \hline
            Block 3 		& \(  2.2 \cdot 10^{-6}\)  &  \checkmark     \\ \hline
            Block 4  		& \( 8.32 \cdot 10^{-7}\)  &  \checkmark   \\ \hline
            ID category 2  	& \( 2.78 \cdot 10^{-7}\)  &  \checkmark   \\ \hline
            ID category 3   & \( 2.04 \cdot 10^{-8}\)  &  \checkmark    \\ \hline
            ID category 4  	& \( 1.20 \cdot 10^{-7}\)  &  \checkmark     \\ \hline
            ID category 5   & \( 2.16 \cdot 10^{-7}\)  &  \checkmark  \\ \hline
        \end{tabular}
        \caption[Friedman test results for accuracy. ]{Friedman test results for accuracy. All groupings pass the test. } \label{tab:accsignifiance}
    \end{center}
\end{table}

\begin{table*}
    \begin{center}
        \begin{tabular}{ | c | c | c | c | c |}
            \hline
            Group & Method 1 & Method 2 & Wilcoxon p-value & Passed? \\ \hline
            \multirow{3}{*}{Overall} 	& ZM & ST & \( 9.56 \cdot 10^{-5} \)  &  \checkmark   \\ \cline{2-5}
            & ZM & PT &  \(  0.11 \cdot 10^{-4} \) & \checkmark \\ \cline{2-5}
            & PT & ST &  \( 9.56 \cdot 10^{-5}\) & \checkmark \\ \hline
            \multirow{3}{*}{Block 1} 	& ZM & ST & \( 9.52 \cdot 10^{-5} \)  &  \checkmark   \\ \cline{2-5}
            & ZM & PT &  \( 0.14 \cdot 10^{-4}\) & \checkmark \\ \cline{2-5}
            & PT & ST &  \( 0.38 \cdot 10^{-4}\) & \checkmark \\ \hline
            \multirow{3}{*}{Block 2} 	& ZM & ST & \( 9.53 \cdot 10^{-5} \)  &  \checkmark   \\ \cline{2-5}
            & ZM & PT &  \( 0.20 \cdot 10^{-4}\) & \checkmark \\ \cline{2-5}
            & PT & ST &  \( 0.23 \cdot 10^{-3}\) & \checkmark \\ \hline
            \multirow{3}{*}{Block 3} 	& ZM & ST & \( 0.11 \cdot 10^{-4} \)  &  \checkmark   \\ \cline{2-5}
            & ZM & PT &  \( 0.20 \cdot 10^{-4}\) & \checkmark \\ \cline{2-5}
            & PT & ST &  \( 0.26 \cdot 10^{-3}\) & \checkmark \\ \hline
            \multirow{3}{*}{Block 4} 	& ZM & ST & \( 0.21 \cdot 10^{-4}\)  &  \checkmark   \\ \cline{2-5}
            & ZM & PT &  \( 0.27 \cdot 10^{-3}\) & \checkmark \\ \cline{2-5}
            & PT & ST &  \( 0.52 \cdot 10^{-4}\) & \checkmark \\ \hline
            \multirow{3}{*}{ID Category 2} 	& ZM & ST & \( 0.19 \cdot 10^{-4}\)  &  \checkmark   \\ \cline{2-5}
            & ZM & PT &  \( 0.14 \cdot 10^{-4}\) & \checkmark \\ \cline{2-5}
            & PT & ST &  \( 0.31 \) & x \\ \hline
            \multirow{3}{*}{ID Category 3} 	& ZM & ST & \( 9.42 \cdot 10^{-5} \)  &  \checkmark   \\  \cline{2-5}
            & ZM & PT &  \( 9.50 \cdot 10^{-5}\) & \checkmark \\ \cline{2-5}
            & PT & ST &  \( 0.42 \cdot 10^{-16}\) & \checkmark \\ \hline
            \multirow{3}{*}{ID Category 4} 	& ZM & ST & \( 9.54 \cdot 10^{-5} \)  &  \checkmark   \\ \cline{2-5} 
            & ZM & PT &  \(  0.11 \cdot 10^{-4}\) & \checkmark \\ \cline{2-5}
            & PT & ST &  \( 0.18 \cdot 10^{-3}\) & \checkmark \\ \hline
            \multirow{3}{*}{ID Category 5} 	& ZM & ST & \( 9.53 \cdot 10^{-5} \)  &  \checkmark   \\ \cline{2-5}
            & ZM & PT &  \( 0.2\) & x \\ \cline{2-5}
            & PT & ST &  \( 0.14 \cdot 10^{-4}\) & \checkmark \\ 
            \hline
        \end{tabular}
        \caption[Results of Wilcoxon signed-rank tests between all combinations of input methods for accuracy. ]{Results of Wilcoxon signed-rank tests between all combinations of input methods. } \label{tab:accposthoc}
    \end{center}
\end{table*}

\subsubsection{User preference}

Table \ref{tab:pref} shows answer results and significance tests for answers in the preference questionaire. During statistical significance testing, results for the overall question did not pass the initial Friedman test and thus no post-hoc tests were conducted. Speed and accuracy passed the Friedman test with \(p =  0.12 \cdot 10^{-4}\) and \( p = 4.48 \cdot 10^{-6}\), respectively. Post-hoc showed no significant differences when comparing ZM vs. ST in terms of speed and PT vs. ST in terms of accuracy (\(p = 0.22  \) and \( p = 1.64 \cdot 10^{-2}\)), but showed significance for ZM vs ST and PT vs. ST in terms of speed ( \(p= 0.20\cdot 10^{-3} \) and \(p= 0.44 \cdot 10^{-4}\)),  as well as ZM vs. ST and ZM vs. PT in terms of accuracy ( \( 0.17 \cdot 10^{-3} \) and  \( 8.09 \cdot 10^{-5} \) ).

\begin{table*}
    \begin{center}
        \begin{tabular}{ | c | c | c | c | }
            \hline
            & ZM & ST & PC \\ \hline
            Speed &  				\(\mu = 2.2, \sigma = 0.61\) &  \(\mu =2.55, \sigma = 0.69\) &  \(\mu = 1.25, \sigma = 0.55\)\\ \hline
            Accuracy &  			\(\mu = 2.8, \sigma = 0.41\) &  \(\mu =1.90, \sigma = 0.64\) &  \(\mu = 1.25, \sigma = 0.44\)\\ \hline
            Overall preference &  	\(\mu = 2.1, \sigma = 0.91\) &  \(\mu =2.20, \sigma = 0.52\) &  \(\mu = 1.70, \sigma = 0.92\)\\ \hline
        \end{tabular}
        \caption[Means and standard deviations for answers given in the preference questionnaire]{Means and standard deviations for answers of different questions in the preference questionnaire. }\label{tab:pref}
    \end{center} 
\end{table*}

\section{Discussion}\label{sec:discussion}

In summary, compared to the smartphone touch input method, Z-Mapping shows better movement times throughout all data groups (overall, by block, by ID), but also worse accuracy throughout all groups. Further, statistical significance could only be shown for accuracy. Direct PC touch input method performs significantly better in terms of movement times than both other input methods. These results align with participant's answers in the preference questionnaire (table \ref{tab:pref}), where users on average reported a slight perceived increase in speed for Z-Mapping over the smartphone touch input case, but almost universally lower perceived accuracy. Further, there is no statistically significant overall preference between input methods in subject's answers. Z-Mapping's failure to show significance for movement times could be because the number of subjects in the study could simply have been too low, or not enough training blocks or non-optimal instructions were provided.  However, movement times grouped by ID show another indication. The gap between Z-Mapping and the smartphone touch method increases from a relative speed up of \(28\%\) at ID category 2 to a \(86\%\) at ID category 5. While further investigations are necessary to confirm it, it may be possible that the chosen output space (PC touch display) is too small relative to the input space to fully show the advantages in movement times of Z-Mapping. Further, data indicates that Z-Mapping suffered disproportionately from below average samples, compared to other input methods. During the experiment, users would often have to make corrective movements after missing a target, resulting in increased movement times. In interviews, participants also often stated that Z-Mapping felt "shaky", which resulted in problems and frustration with target selection (confirmed by accuracy results). This hints at a need to improve the final target selection step of Z-Mapping. Overall, the method needs further work.

\section{Future Work}\label{sec:futurework}

Z-Mapping, but also Z-Scaling, have not been evaluated for larger scale output spaces in this thesis. The chosen PC touch display is still small enough that users can reach targets by physically touching their on-screen locations. Of interest then are environments where users are unable to physically reach targets, or where doing such would imply very large movement times (e.g. walking to the other end of the room). One such example are VR spaces in general, where users may locally only be able to physically reach a very small fraction of the overall output space. Investigations towards that end could also confirm or deny whether the general trend of Z-Mapping performing better at higher IDs continues at higher distances.

One major problem of Z-Mapping as determined during the user study is the final phase of target acquisition when users try to click on a target. As the environment keeps moving on downwards finger movements, and users don't typically move their fingers in perfect lines that are orthogonal towards the target, user's experience a slight shaking of the environment that results in missed clicks. It may be possible to work with the user's tracked finger trajectory to determine the point at which they begin trying to click targets, and from that point switch to a different input mapping scheme that is optimized for hitting the target below the finger. Another possible point of investigation would be techniques that detach the moving of the environment from the final target selection. In Z-mapping during the evaluated user study, the environment always moved whenever the user's fingers moved. Instead, it might be possible to deactivate this movement, e.g. via secondary controls activated by a user's unused hand. The same concepts could be used give the user a way to quickly adjust their positioning when they miss a target on their first try, which in the user study required them to touch the smartphone screen to disable environment movement and move back towards the middle in order to try again. 

\section{Summary}

This thesis presented  C-D ratio mappings for commonplace as well as novel input techniques and implemented the presented concepts in two distinct multi-display environments. Z-Mapping in particular was then evaluated in a user study by comparing it against touch based input techniques as they are commonly used in smartphone applications. As expected, it performs worse overall compared to the direct PC-touch screen input method that was used as a baseline comparison. While in the collected data Z-Mapping was shown to perform better in average movement times compared to the smartphone touch input method, the difference was not found to be statistically significant. In addition, the study showed problems with accuracy in target selection. In informal interviews with the study subjects, this has been determined to likely be because of target acquisition problems due to the high sensitivity of the input method in general and a lack of possibility for users to intuitively adjust the final positioning if they miss the target on the first try. Finally, possible approaches for future work in order to improve various aspects of the concepts shown in this thesis were presented.

\bibliographystyle{abbrv}
\bibliography{template}
\end{document}